\documentclass[cameraready]{Interspeech}

\title{Beyond Speaker Independence: Evaluating Cross-Lingual Acoustic-to-Articulatory Inversion Across Finnish and Russian}

\author{Ruchi Pandey and Tomi H. Kinnunen}

\address{
   University of Eastern Finland, Finland
}

\email{ruchi.pandey@uef.fi, tomi.kinnunen@uef.fi}

\keywords{Acoustic-to-articulatory inversion, Electromagnetic articulometry, Self-supervised learning, FROST-EMA}

\usepackage{comment}
\usepackage{makecell}
\usepackage{multirow}
\usepackage{booktabs}
\usepackage[table]{xcolor}

\begin{document}

\maketitle
\begin{abstract}
Acoustic-to-articulatory inversion (AAI) remains challenging under domain shifts where changes in speaker attributes and cross-language conditions often degrade performance. We conduct a systematic evaluation under such shifts and establish baseline benchmarks on FROST-EMA, a Finnish–Russian bilingual EMA corpus. FROST-EMA addresses the English bias and limited speaker diversity of existing resources. We benchmark (i) articulatory targets (raw EMA coordinates vs tract variables), (ii) acoustic front-ends (MFCC vs SSL features), and (iii) inversion back-ends (BiLSTM vs a lightweight attention-based sequence model). We further define evaluation protocols for cross-gender transfer (within language) and cross-language transfer (within gender). The results indicate that cross-gender mismatch introduces moderate Pearson correlation declines (approx. 0.05–0.10) relative to the in-domain baseline, whereas cross-language mismatch causes larger drops (approx 0.10- 0.20).



\end{abstract}

\section{Introduction}
\label{sec:intro}
Acoustic-to-articulatory inversion (AAI) refers to estimating time-varying vocal-tract movements from the acoustic speech signal \cite{atal1978inversion,browman1986towards}. AAI is an ill-posed problem, as similar acoustic realizations can arise from different articulatory configurations \cite{lindblom1977formant}. This one-to-many mapping from acoustics to articulation makes AAI a highly non-linear inversion problem \cite{lindblom1977formant,qin2007empirical}. Despite these challenges, reliable estimation of articulatory trajectories has many applications (e.g., speech recognition \cite{mitra2010articulatory,ghosh2011automatic}, speech synthesis \cite{6289354}, and pronunciation training \cite{suemitsu2015real}). As in other speech tasks, deep neural networks (DNNs) have become the dominant approach for AAI. DNNs model coarticulation effects and long-range temporal dependencies more effectively 
compared to shallow statistical methods \cite{zhang2008acoustic,ghosh2013smoothing}. Architectures such as bidirectional long short-term Memory (BiLSTMs) have emerged as strong baselines \cite{zhu2015articulatory,illa2018low}, while attention-based models further improve temporal flexibility \cite{shahrebabaki2021raw,udupa2021estimating,chung2024speaker}.

A key bottleneck in AAI research is the scarcity of suitable electromagnetic articulography (EMA) corpora. Most publicly available datasets are small, predominantly English, and offer limited diversity in speakers, languages, and recording conditions (e.g., MOCHA-TIMIT and mngu0) \cite{wrench1999mocha,richmond2011announcing}. This limits the possibilities for addressing speaker- and language- independent AAI. Prior cross-linguistic and cross-accent studies consistently report performance degradation when training and test conditions differ \cite{wieling2017analysis,illa2022impact,yan2023combining}. These findings highlight the need for evaluation protocols that isolate specific sources of domain shift, such as speaker identity, gender, and language. Given these data constraints, self-supervised learning (SSL) 
representations provide a useful acoustic front-end for 
AAI on under-represented languages. Trained on large unlabeled speech corpora, SSL front-ends can extract rich acoustic representations without requiring language-specific articulatory data, and recent studies suggest they encode information closely related to articulatory 
structure~\cite{hao2024exploring,cho2024self,bandekar2025enhancing}.

\begin{figure}[t]
  \centering
  \includegraphics[width=\linewidth]{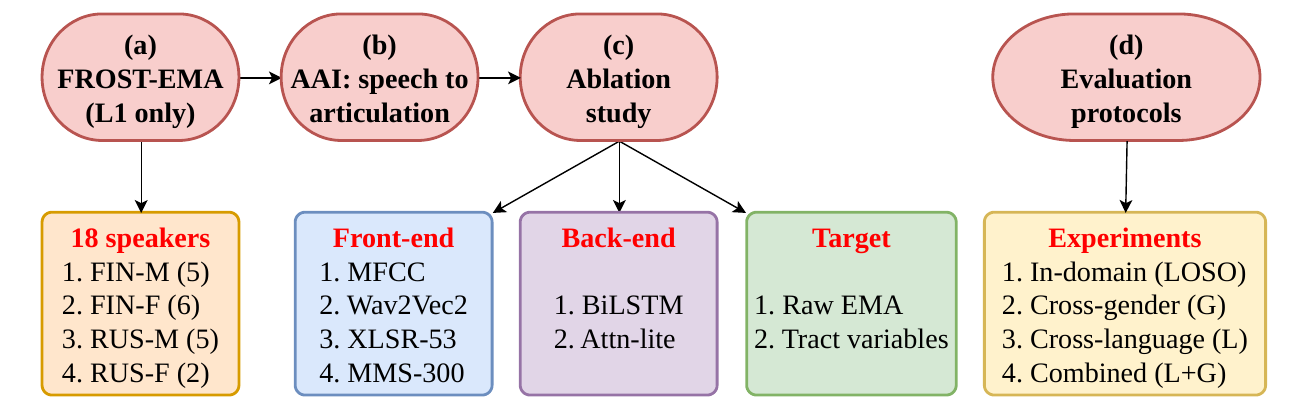}
  \caption{Overview of experimental design for AAI: (a) FROST-EMA speaker groups, (b) AAI, (c) Ablation study, and (d) Evaluation protocols.}
  \label{fig:blk_diagram}
  \vspace{-0.7cm}
\end{figure}

Despite these advances, cross-domain (language and gender) speaker-independent AAI remains largely underexplored. The two closest prior works evaluated cross-language AAI between English--Dutch \cite{wieling2017analysis} and English--Japanese \cite{yan2023combining}, but neither isolated gender as a domain shift factor, employed SSL front-ends, nor conducted all evaluations within a single recording protocol. The preliminary work \cite{pollio2026presence} applied linear probing to FROST-EMA to assess layer-wise articulatory encoding in SSL models, but did not train nonlinear inversion models or evaluate cross-gender and cross-language transfer. In this work, we benchmark AAI on FROST-EMA \cite{hopponen2025frost}, a bilingual Finnish--Russian EMA corpus of 18 speakers recorded under a uniform protocol. Finnish and Russian are typologically distant from the Germanic and Japonic families in prior work, introducing contrasts such as vowel harmony and pervasive palatalization that broaden the empirical basis for assessing inversion robustness. The uniform protocol enables clean isolation of gender and language as independent domain shift factors. We conduct a systematic ablation over acoustic front-ends, articulatory target representations, and back-end architectures (detailed in Section~\ref{sec:AAI_model}). The experimental design is shown in Figure~\ref{fig:blk_diagram}. The main contributions are as follows:
\begin{itemize} 
    \item We introduce standardized AAI benchmarking and preprocessing for FROST-EMA.
    \item For the first time in cross-lingual AAI, we define controlled evaluation protocols that isolate cross-gender (within-language) and cross-language (within-gender) shifts as independent factors.
    \item First systematic ablation across acoustic front-ends, articulatory target representations, and back-end architectures within a single bilingual corpus and evaluation framework.
\end{itemize}

\section{Dataset }
FROST-EMA is a recently collected electromagnetic articulography corpus designed to address the limited coverage and English bias of existing parallel acoustic--EMA resources~\cite{hopponen2025frost}. It contains recordings from 18 bilingual speakers (11 native Finnish, 7 native Russian; 8 female, 10 male) under three controlled speaking conditions: native language (L1), second language (L2), and imitated L2 (fake foreign accent). These speakers form four language-gender groups: FIN-M (5), FIN-F (6), RUS-M (5), and RUS-F (2). Unlike the two widely adopted corpora, MOCHA-TIMIT \cite{wrench1999mocha} and mngu0 \cite{richmond2011announcing}, FROST-EMA provides non-English data, explicit accent control, a much larger speaker pool, and high-rate articulatory trajectories captured at 1250 Hz with a state-of-the-art AG501 articulograph. Articulatory data were captured using five midsagittal sensors, with no jaw sensor: upper lip (UL), lower lip (LL), tongue tip (TT), tongue blade (TB), and tongue dorsum (TD). For each sensor, the X (back--front) and Z (up--down) coordinates were extracted, yielding 10 continuous articulatory features per frame \cite{pollio2026presence}.
In this work, EMA segments are automatically screened for reliability, preprocessed with a standardized pipeline explained later. To isolate anatomical and phonological factors from proficiency-related variability, all the AAI models are trained and evaluated exclusively on L1 productions under controlled cross-gender and cross-language splits. 
\section{Proposed AAI Model}
\label{sec:AAI_model}
\subsection{Preprocessing}
We extract articulatory trajectories from head-movement-corrected AG501 positional recordings sampled at 1250\,Hz. Channel-to-sensor mapping follows the default FROST-EMA 
protocol~\cite{hopponen2025frost}, with speaker-specific overrides where sensor reassignment is documented. We preprocess each of the 10 articulatory channels independently through four stages: (1) brief dropouts are replaced via linear interpolation, while segments with extended signal loss are excluded; (2) a 6th-order 
zero-phase Butterworth low-pass filter at 20\,Hz suppresses 
measurement noise and provides anti-aliasing; (3) filtered signal is decimated from 1250\,Hz to 50\,Hz using polyphase resampling to match the acoustic frame rate; and (4) per-utterance, per-channel z-score normalization is applied to reduce anatomical offsets and standardize amplitudes.

\subsection{Articulatory Target Representations}
We evaluate AAI models under two articulatory target representations: Raw EMA and tract variables (TVs) \cite{boe1992geometric}. 
For Raw EMA, the model directly predicts the z-scored midsagittal sensor coordinates: UL$_x$, UL$_z$, LL$_x$, LL$_z$, TT$_x$, TT$_z$, TB$_x$, TB$_z$, TD$_x$, TD$_z$ ($D{=}10$). For TVs, we derive a compact five-dimensional constriction-based representation from the physical (non-normalized) EMA
coordinates~\cite{mcgowan1994recovering,sivaraman2019unsupervised}. TVs encode functionally relevant articulatory parameters and prior work suggests TVs are more speaker-invariant than raw sensor positions \cite{wieling2017analysis}, making them a natural comparison point for speaker-independent AAI. Following established definitions \cite{sivaraman2019unsupervised,wieling2017analysis}, the five TVs are computed from the FROST-EMA sensors. Lip aperture (LA) is the Euclidean distance between UL and LL. Standard lip protrusion (LP) is LL X-displacement from its utterance-level median. For tongue sensors (TT, TB, TD), constriction location (CL) is defined as median-relative X/back-front displacement, yielding three variables: TTCL, TBCL, and TDCL. Per-utterance z-normalization is applied independently to each of the five TV dimensions, standardizing scale across utterances while preserving the geometric relationships used in TV computation. 
We omit constriction degree (CD) variables because the required palate-trace references are unavailable in FROST-EMA.
\subsection{Acoustic-to-Articulatory Inversion Pipeline}
The AAI pipeline consists of an acoustic front-end and an inversion back-end (see Figure~\ref{fig:blk_diagram}).  We consider two classes of front-ends: a DSP-based spectral 
feature baseline that offers simplicity and low computational cost, and modern self-supervised representations that capture richer speech structure at the expense of higher dimensionality. As a DSP-based baseline, we extract 40-dimensional MFCCs using a 25\,ms window, 10\,ms frame shift, and 40 mel filters spanning 20--8000\,Hz. In addition, we extract frozen self-supervised representations from three pretrained models: Wav2Vec~2.0 Base (768-dim)~\cite{baevski2020wav2vec}, XLSR-53 Large (1024-dim)~\cite{babu2021xls}, and MMS-300m (1024- dim)~\cite{pratap2024scaling}. We use the final encoder 
layer, following prior AAI work that found upper layers encode more articulatory-relevant information \cite{cho2024self}. No fine-tuning is applied to any SSL model. None of these models include FROST-EMA speakers in their pretraining data, ensuring no speaker overlap between front-end training and our evaluation. For training, the aligned feature--articulatory pairs are segmented into non-overlapping windows of 100 frames ($\approx$\,2\,s). Utterances shorter than 100 frames are padded by repeating the final frame.

We consider two back-end architectures based on earlier AAI 
studies~\cite{zhu2015articulatory,illa2018low,udupa2021estimating,chung2024speaker}. The first is a BiLSTM with two layers and a hidden size of 256 per direction, followed by a two-layer MLP that produces per-frame articulatory estimates. The second is a lightweight Transformer encoder (hereafter `Attn-lite') with 4 self-attention layers using pre-layer normalization, 4 attention heads, an embedding dimension of 256, and a feedforward dimension of 512. Input features are projected to the embedding space, augmented with sinusoidal positional encoding, and processed by the self-attention layers. A two-layer MLP head produces the final predictions. Both models minimize mean squared error (MSE) between predicted and reference trajectories using Adam with a learning rate of $1\times10^{-3}$ and a batch size of 8. Training runs for up to 50 epochs with early stopping (patience\,=\,8) based on validation loss. Validation data are drawn from a random 10\% of held-out training utterances. During training, a random segment is cropped from each utterance per epoch; during evaluation, the center segment is used. Following standard AAI evaluation practice \cite{weise2025towards,wieling2017analysis}, we 
report Pearson correlation ($r$) between predicted and 
reference trajectories computed per articulatory dimension. 
\section{Experiments and Results}
\subsection{In-domain speaker-independent baselines}
\begin{figure}[]
  \centering
  \includegraphics[width=0.93\linewidth]{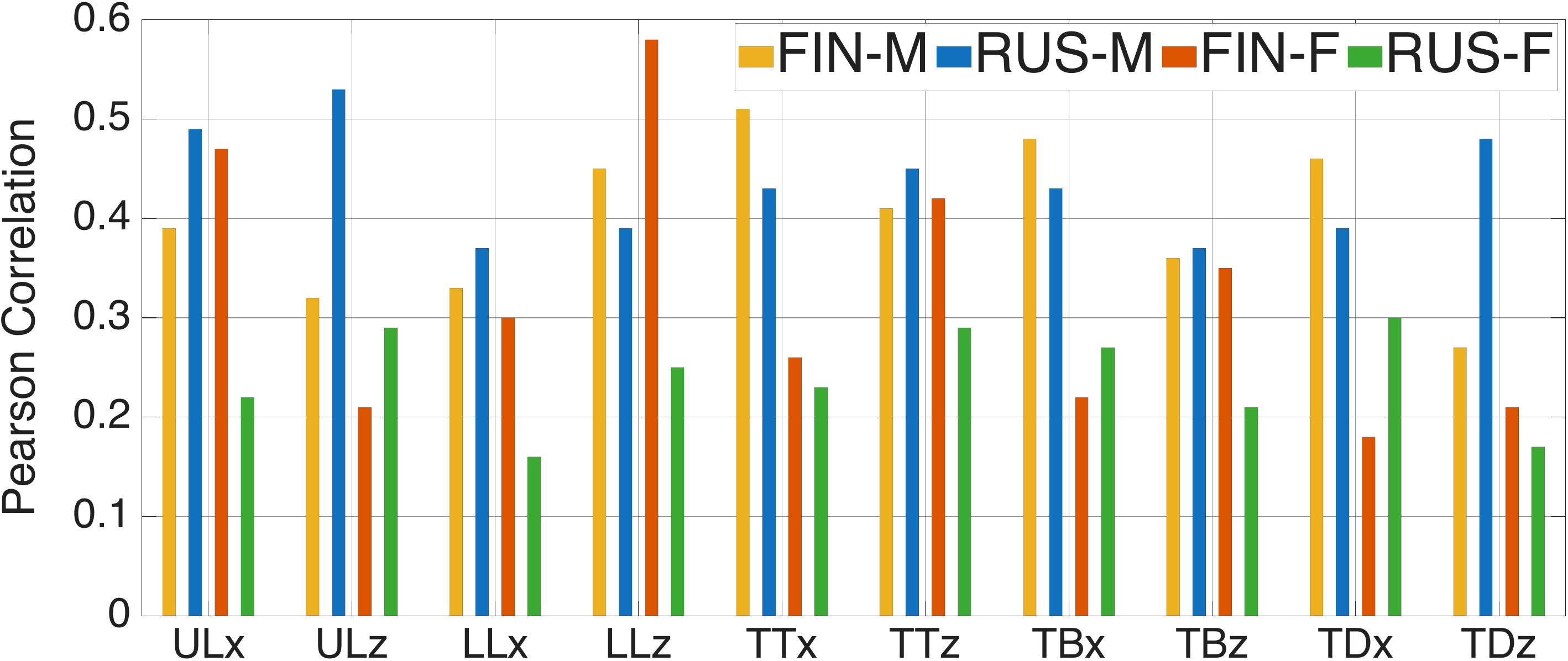}
  \caption{Per-channel Pearson correlation ($r$) for raw EMA targets across four language–gender groups (FIN-M, FIN-F, RUS-M, RUS-F) under in-domain LOSO evaluation using Wav2Vec 2.0 with BiLSTM} 
  \label{fig:barplot_ema}
   \vspace{-0.3cm}
\end{figure}
\begin{figure}[]
  \centering
  \includegraphics[width=0.92\linewidth]{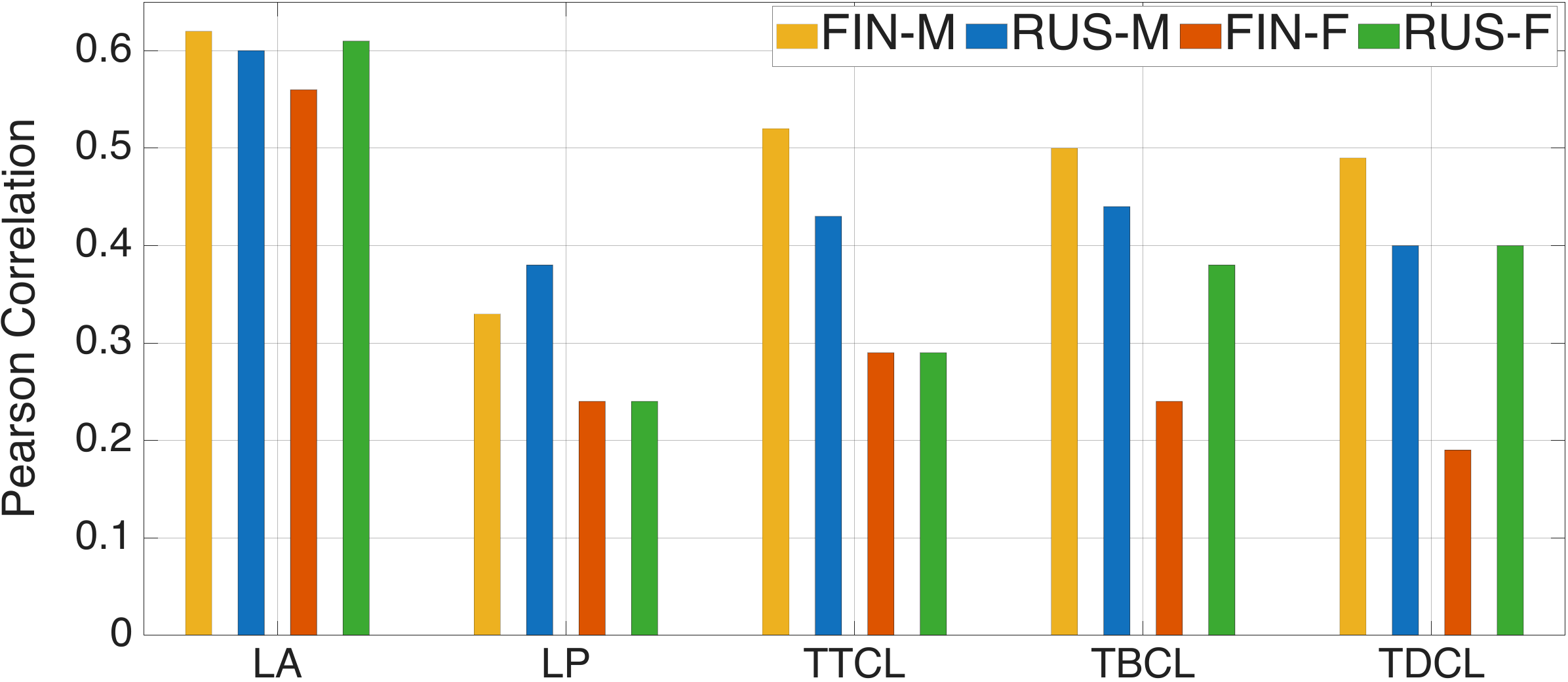}
  \caption{Per-channel Pearson correlation ($r$) for tract variable targets across four language–gender groups under in-domain LOSO evaluation using Wav2Vec 2.0 with BiLSTM.}
  \label{fig:barplot_tv}
  \vspace{-0.6cm}
\end{figure}
We first establish in-domain baselines using leave-one-speaker-out (LOSO) evaluation within each language–gender group (FIN-M, FIN-F, RUS-M, and RUS-F). LOSO ensures that the test speaker is never seen during training, providing a rigorous measure of cross-speaker generalization. Validation data are drawn from utterance-level splits of training speakers; test speakers are fully held out. Figures \ref{fig:barplot_ema} and \ref{fig:barplot_tv} show per-channel Pearson correlations averaged across all four groups for raw EMA and TV targets, respectively, using Wav2Vec 2.0 front-end with BiLSTM. For raw EMA targets (Figure \ref{fig:barplot_ema}), tongue sensors (TT, TB, TD) consistently outperform the lip sensors (UL, LL), with vertical (Z-axis) coordinates predicted better than horizontal (X-axis). This reflects stronger acoustic coupling of vertical tongue displacement to formant structure. ULz remains the most challenging channel across all groups. 

Figure~\ref{fig:barplot_tv} indicates that LA yields moderate correlations, while LP is systematically the weakest TV across all four groups. This suggests that lip protrusion dynamics are poorly coupled to the acoustic signal relative to aperture-based and tongue-based constrictions. Among tongue TVs, all constriction location variables (TTCL, TBCL, TDCL) display broadly comparable correlations. The five-dimensional TV representation retains the core articulatory contrasts while offering a more interpretable decomposition of lip versus tongue constriction behavior. This hierarchy is stable in this corpus, suggesting that CL is more acoustically recoverable than LP, though broader generality requires more languages. Overall, TV correlation trends are broadly comparable with EMA targets.
\subsection{Cross-gender transfer within language}
Table \ref{tab:cross_gender_SI} reports per-articulator correlations for cross-gender transfer (language held constant), isolating gender-related anatomical variability. Interestingly, Table \ref{tab:cross_gender_SI} indicates that cross-gender degradation depends on the direction of transfer. For Finnish, FIN-F$\rightarrow$FIN-M consistently outperforms FIN-M$\rightarrow$FIN-F, particularly for tongue sensors (TT: 0.46 vs. 0.34; TD: 0.33 vs. 0.23). This may partly reflect the larger FIN-F pool (6 vs. 5). For Russian, the directional gap is smaller despite the severely limited RUS-F pool (2 speakers). This may indicate greater variability in RUS-M, but RUS-F has only two speakers.
\newcommand{\cellr}[2]{\makecell{#1\\{\footnotesize(#2)}}} 
\begin{table}[h]
\centering
\caption{Cross-gender speaker-independent transfer (language held constant, L1 only). Per-articulator Pearson $r$ using Wav2Vec 2.0 with BiLSTM on raw EMA targets.}
\label{tab:cross_gender_SI}
\setlength{\tabcolsep}{3.5pt} 
\renewcommand{\arraystretch}{1.00} 
\begin{tabular}{lllllll}
\toprule
Train & Test  & LL & UL & TT & TB & TD \\ \hline
FIN-M & FIN-F & \cellr{0.46}{0.090} & \cellr{0.35}{0.046} & \cellr{0.34}{0.085} & \cellr{0.31}{0.087} & \cellr{0.23}{0.008} \\ \hline
FIN-F & FIN-M & \cellr{0.43}{0.060} & \cellr{0.38}{0.057} & \cellr{0.46}{0.065} & \cellr{0.42}{0.065} & \cellr{0.33}{0.066} \\ \hline
RUS-M & RUS-F & \cellr{0.41}{0.052} & \cellr{0.46}{0.035} & \cellr{0.42}{0.026} & \cellr{0.44}{0.057} & \cellr{0.43}{0.052} \\ \hline
RUS-F & RUS-M & \cellr{0.35}{0.066} & \cellr{0.48}{0.036} & \cellr{0.41}{0.043} & \cellr{0.42}{0.050} & \cellr{0.39}{0.036} \\ \bottomrule
\end{tabular}
\end{table}

\begin{figure}[h!]
  \centering
  \includegraphics[width=\linewidth]{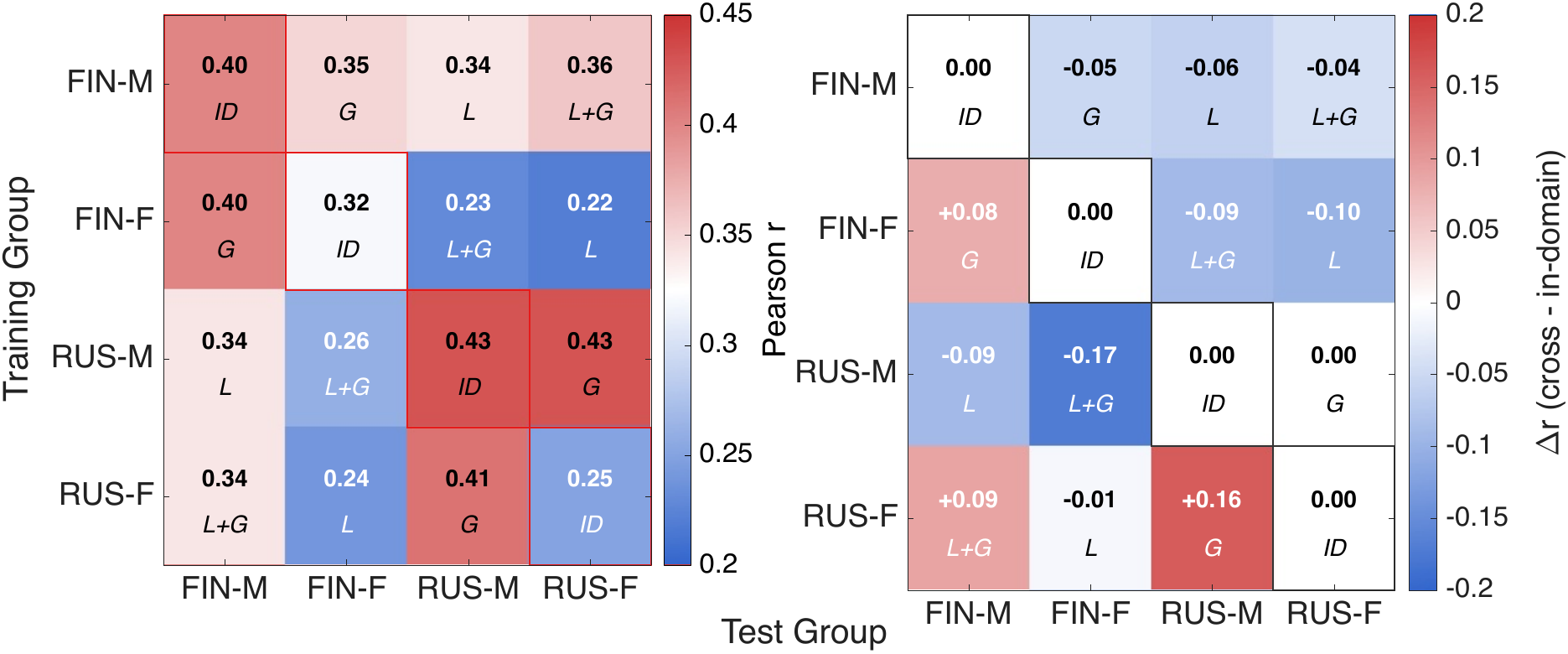}
  \caption{ Cross-domain AAI transfer (Wav2Vec 2.0, BiLSTM, raw EMA). (a) Absolute Pearson $r$ for all training–test group pairs (left). (b) Performance drop ($\Delta r$) relative to in-domain diagonal baselines (right). Cell labels indicate shift type: ID (in-domain), G (gender), L (language), L+G (combined).}
  \label{fig:Heatmap}
 \vspace{-0.6cm}
\end{figure}
\begin{figure}[h]
  \centering
  \includegraphics[width=0.95\linewidth]{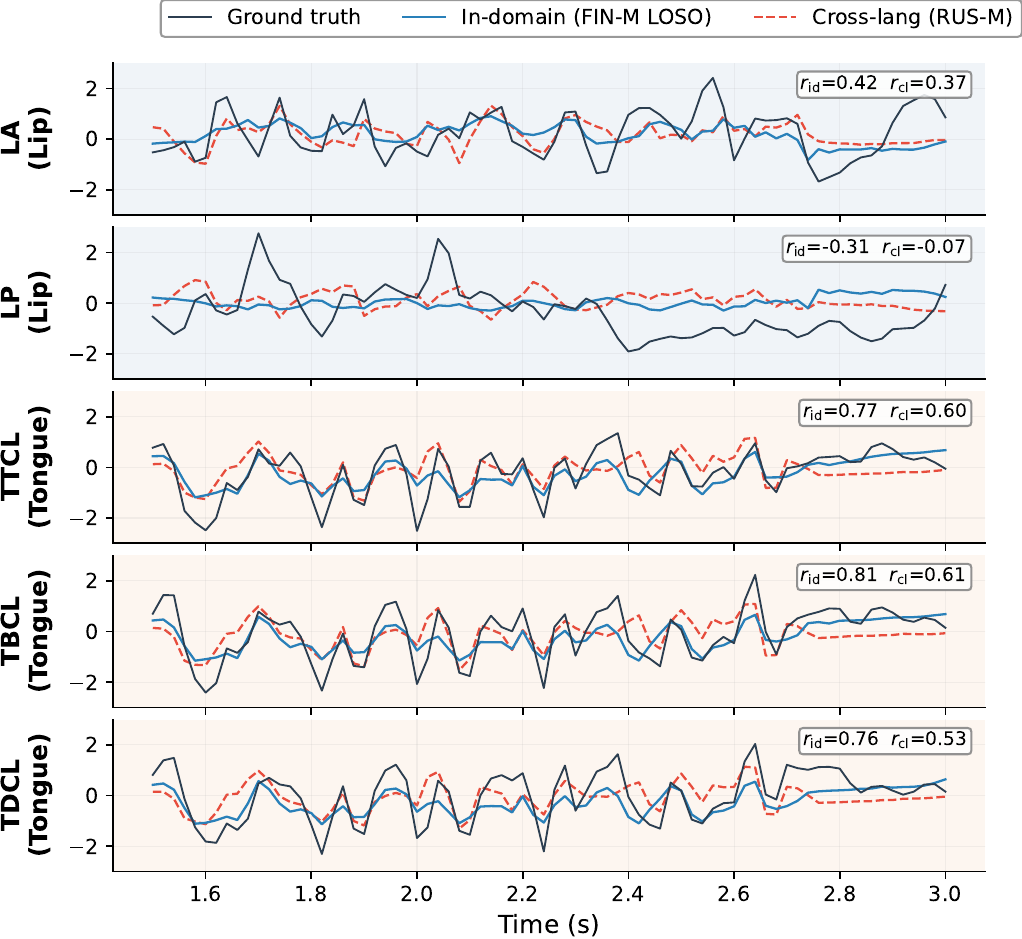}
  \caption{Predicted vs. ground-truth tract variable trajectories for a representative FIN-M test utterance under in-domain (FIN-M LOSO) and cross-language (RUS-M $\rightarrow$ FIN-M) conditions, using Wav2Vec 2.0 with BiLSTM. }
  \label{fig:tv_plot}
\vspace{-0.6cm}
\end{figure}
Lip sensors exhibit a language-dependent pattern. In Finnish, LL outperforms UL, whereas in Russian, UL outperforms LL. This reversal reflects acoustic predictability, not necessarily UL mobility, and may relate to language-specific rounding/protrusion strategies \cite{pollio2026presence}. TD is the most affected articulator under Finnish gender transfer (dropping to $r=0.23$ for FIN-M$\rightarrow$FIN-F), while remaining robust in Russian ($r\geq 0.39$). Since TD captures tongue dorsum positioning relevant to 
velar constrictions, this disparity may reflect the 
broader palatalization contrasts in Russian, which engage 
the tongue dorsum more extensively than Finnish velar 
consonants. Overall, cross-gender transfer introduces a systematic correlation drop of approximately $0.05$–$0.10$ relative to in-domain results.
\definecolor{sepgray}{HTML}{C0C0C0}
\begin{table}[t]
\centering
\caption{Ablation over front-ends, targets, and back-ends. Pearson $r$
(std.\,dev.) on three conditions anchored to FIN-M: in-domain (ID) LOSO,
cross-gender (FIN-M$\to$FIN-F), and cross-language (FIN-M$\to$RUS-M).}
\label{tab:tab_ablation}
\setlength{\tabcolsep}{1.3pt}
\renewcommand{\arraystretch}{0.80}
\small
\newcommand{\rv}[2]{\makecell{#1\\[-0.5pt]{\scriptsize(#2)}}}
\newcommand{\grayrule}{%
  \arrayrulecolor{sepgray}\cmidrule(lr){3-8}\arrayrulecolor{black}}
\begin{tabular}{@{}ll ccc ccc@{}}
\toprule
\textbf{Target} & \textbf{Front-end}
& \multicolumn{3}{c}{\textbf{BiLSTM}} 
& \multicolumn{3}{c}{\textbf{Attn-lite}} \\
\cmidrule(lr){3-5} \cmidrule(lr){6-8}
& 
& \multicolumn{1}{c}{\textbf{LOSO}} 
& \multicolumn{2}{c}{\textbf{Cross-domain}} 
& \multicolumn{1}{c}{\textbf{LOSO}} 
& \multicolumn{2}{c}{\textbf{Cross-domain}} \\
\cmidrule(lr){3-3} \cmidrule(lr){4-5} 
\cmidrule(lr){6-6} \cmidrule(lr){7-8}
& 
& ID & G & L 
& ID & G & L \\
\midrule
\multirow{10}{*}{\textbf{EMA}}
  & MFCC    & \rv{0.30}{0.028} & \rv{0.27}{0.056} & \rv{0.32}{0.029}
            & \rv{0.24}{0.018} & \rv{0.16}{0.047} & \rv{0.20}{0.024} \\
\grayrule
  & Wav2Vec & \rv{0.40}{0.026} & \rv{0.35}{0.062} & \rv{0.35}{0.037}
            & \rv{0.34}{0.022} & \rv{0.30}{0.061} & \rv{0.31}{0.031} \\
\grayrule
  & XLSR-53 & \rv{0.31}{0.019} & \rv{0.24}{0.066} & \rv{0.20}{0.037}
            & \rv{0.27}{0.027} & \rv{0.25}{0.069} & \rv{0.21}{0.054} \\
\grayrule
  & MMS     & \rv{0.41}{0.030} & \rv{0.34}{0.051} & \rv{0.30}{0.027}
            & \rv{0.37}{0.027} & \rv{0.33}{0.072} & \rv{0.32}{0.034} \\
\midrule
\multirow{10}{*}{\textbf{TV}}
  & MFCC    & \rv{0.42}{0.026} & \rv{0.28}{0.075} & \rv{0.36}{0.036}
            & \rv{0.29}{0.026} & \rv{0.19}{0.057} & \rv{0.26}{0.038} \\
\grayrule
  & Wav2Vec & \rv{0.49}{0.019} & \rv{0.34}{0.070} & \rv{0.39}{0.031}
            & \rv{0.42}{0.016} & \rv{0.30}{0.081} & \rv{0.35}{0.042} \\
\grayrule
  & XLSR-53 & \rv{0.40}{0.020} & \rv{0.25}{0.085} & \rv{0.24}{0.062}
            & \rv{0.36}{0.025} & \rv{0.27}{0.084} & \rv{0.28}{0.058} \\
\grayrule
  & MMS     & \rv{0.49}{0.013} & \rv{0.35}{0.069} & \rv{0.36}{0.041}
            & \rv{0.44}{0.027} & \rv{0.30}{0.064} & \rv{0.29}{0.045} \\
\bottomrule
\end{tabular}
\vspace{-0.5cm}
\end{table}
\subsection{Cross-language transfer within gender}
Figure \ref{fig:Heatmap} summarizes all training–test combinations. The heatmap reveals clear directional asymmetries: FIN-M$\rightarrow$RUS-M and RUS-M$\rightarrow$FIN-M yield identical overall correlations ($r=0.34$). Note that FIN-F$\rightarrow$RUS-F ($r = 0.22$) underperforms RUS-F$\rightarrow$FIN-F ($r=0.25$), though both values should be interpreted with caution given the 
limited RUS-F pool (2 speakers). Combined language-plus-gender shifts (L+G) display the largest degradation, confirming that anatomical and phonological shifts compound rather than cancel. Figure \ref{fig:Heatmap} shows RUS-M generalizes well to both FIN-M ($r=0.34$) and RUS-F ($r=0.43$), suggesting that Russian male training pool captures sufficient articulatory diversity for moderate cross-domain transfer consistent with the cross-gender findings.

To examine which articulatory dimensions are most affected by language mismatch, we analyze TV predictions rather than raw EMA coordinates. Figure \ref{fig:tv_plot} depicts ground-truth, in-domain (FIN-M LOSO), and cross-language (RUS-M $\rightarrow$ FIN-M) TV predictions for a representative test utterance. LA predictions remain closely aligned across both conditions ($r_{id} = 0.38, r_{cl}=0.36$). 
LP yields weak correlations under both conditions ($r_{id} = -0.18, r_{cl} = 0.00$), confirming that lip protrusion is poorly recoverable from acoustics even within matched conditions. All three tongue TVs achieve substantially higher in-domain correlations than lip TVs. Under cross-language transfer, the tongue CL predictions visibly diverge from the ground truth during rapid articulatory transitions $\Delta r \approx 0.04 - 0.07$, a larger degradation than for LA. This is  consistent with tongue CL variables being more language-specific, reflecting differences in consonant place contrasts and palatalization strategies between Finnish and Russian. The trajectory-level visualization (see Fig. \ref{fig:tv_plot}) provides diagnostic evidence beyond aggregate correlation statistics, revealing that cross-language degradation concentrates at phonemically active segments rather than occurring uniformly across the utterance.
\subsection{Ablation Study}
Table \ref{tab:tab_ablation} compares all combinations of acoustic front-end, targets, and inversion back-ends on three representative conditions anchored to FIN-M (chosen for its balanced pool size and availability of both cross-gender and cross-language targets). We conclude the following: 
\begin{itemize}
    \item \textbf{Front-ends:} SSL representations consistently outperform MFCCs, with Wav2Vec 2.0 and MMS-300m achieving the highest correlations across all conditions and both back-ends. XLSR-53 underperforms despite its multilingual pretraining, possibly due to limited Finnish/Russian exposure in its training data. 
    The relative ranking of front-ends is preserved under domain shift, suggesting that articulatory information encoded in Wav2Vec~2.0 and MMS representations is partially retained across speakers and languages.
    \item \textbf{Back-ends:} BiLSTM consistently matches or outperforms Attn-lite across all front-end and target combinations. This suggests that BiLSTM's recurrent inductive bias benefits articulatory trajectory modeling, especially with lower-dimensional features. The MFCC-Attn-lite combination yields the weakest results overall ($r=0.24$-$0.29$), indicating that lightweight attention model requires richer input representations, though limited training data may also contribute.
    \item \textbf{Target representation:} TVs and EMA achieve comparable aggregate accuracy, despite the substantial difference in output dimensionality. The main advantage of TVs is its semantically interpretable dimensions that enable diagnostic analysis of which articulatory aspects transfer across domains, an analytical capability absent from raw EMA evaluation.
\end{itemize}
\section{Conclusion}

We presented the first systematic, speaker-independent AAI benchmarks on non-English bilingual FROST-EMA corpus of 18 Finnish-Russian speakers. The key findings are as follows:
\begin{itemize}
    \item Language mismatch ($\Delta r \approx 0.10$–$0.20$) degrades AAI more than gender mismatch ($\Delta r \approx 0.05$–$0.10$) and combined shifts (L+G) produce the largest drops. This suggests that anatomical and phonological mismatches compound rather than cancel. Prior cross-language studies \cite{wieling2017analysis,yan2023combining} did not quantify these asymmetries separately, as language and recording-protocol effects were confounded.

    \item TVs and raw EMA achieve comparable aggregate accuracy, but TVs offer a diagnostic advantage. Lip TVs remain robust across languages, while tongue CL variables degrade substantially, suggesting that constriction location is more language-specific due to differences in place contrasts and palatalization between Finnish and Russian.

    \item Wav2Vec 2.0 and MMS-300m features consistently outperformed MFCCs under both matched and mismatched conditions: the front-end ranking is preserved under domain shift, indicating that articulatory information in these representations partially transfers across speakers and languages.

    \item BiLSTM outperformed lightweight attention model across all configurations suggesting that recurrent inductive bias is beneficial under low-resource EMA training conditions.

   \item Future work will extend evaluation to L2 and accented speech conditions available in FROST-EMA, investigate speaker adaptation strategies.
\end{itemize}



\section{Generative AI Use Disclosure}
Generative AI tools were used for minor editing and polishing of the manuscript. Authors carefully reviewed and edited the content and take full responsibility for the publication.


\bibliographystyle{IEEEtran}
\bibliography{mybib}

\clearpage

\setcounter{section}{0}
\setcounter{subsection}{0}
\setcounter{figure}{0}
\setcounter{table}{0}
\renewcommand{\thesection}{S\arabic{section}}
\renewcommand{\thefigure}{S\arabic{figure}}
\renewcommand{\thetable}{S\arabic{table}}

\twocolumn[{%
  \begin{minipage}{\textwidth}
    \centering
    {\large\bfseries Supplementary Material With Additional Results\par}
    \vspace{2em}
    \begin{minipage}{\textwidth}
      \noindent
      These supplementary pages present additional experimental results for the above study titled “Beyond Speaker Independence: Evaluating Cross-Lingual Acoustic-to-Articulatory Inversion Across Finnish and Russian”. Due to an unnoticed bug, the EMA data was erroneously low-pass filtered at 4 Hz prior to decimating it to match SSL frame-rate. The original results therefore represent AAI performance using lower-rate, more strongly smoothed articulatory trajectories. To complement these findings, we repeat the experiments using the 20 Hz low-pass cutoff and resampled the EMA trajectories to 50 Hz. The original experimental configurations, including acoustic front-ends, model architectures, training and evaluation protocols, are retained. All experiments are recomputed from scratch without reusing cached features or data. Since the two configurations also differ in effective filtering bandwidth and the temporal duration represented by fixed-length segments, the comparison reflects the overall preprocessing configurations rather than an isolated sampling-rate ablation.
    \end{minipage}
    \vspace{1.2em}
  \end{minipage}
}]

\section{Additional Results at 50 Hz}
\subsection{In-domain speaker-independent baselines}
\begin{figure}[h]
  \centering
  \includegraphics[width=0.93\linewidth]{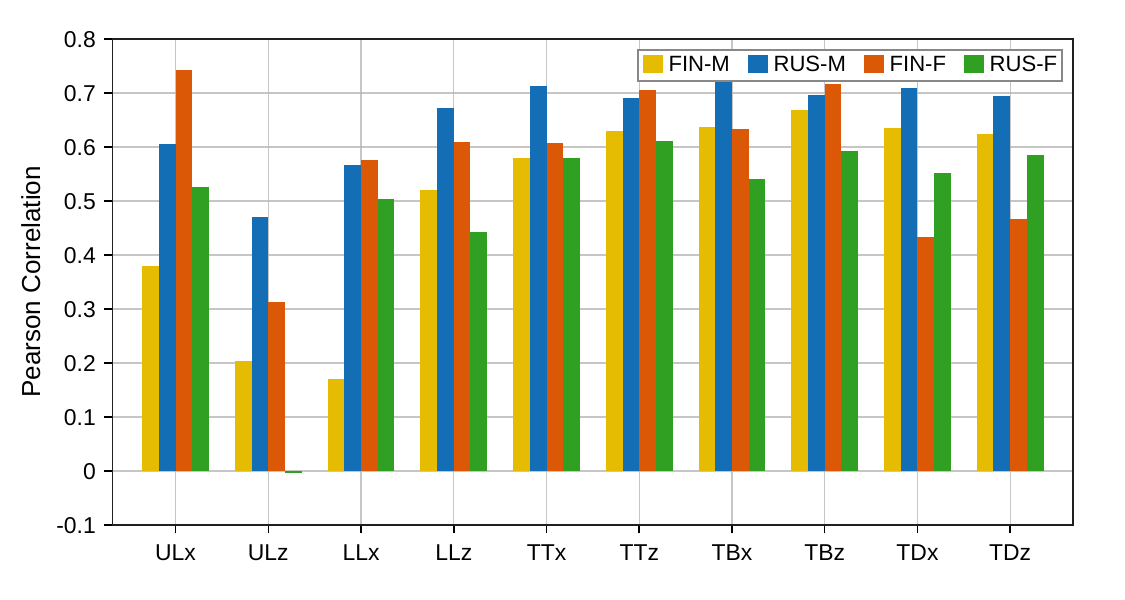}
  \caption{Per-channel Pearson correlation ($r$) for raw EMA targets across four language–gender groups (FIN-M, FIN-F, RUS-M, RUS-F) under in-domain LOSO evaluation using Wav2Vec 2.0 with BiLSTM} 
  \label{fig:barplot_ema_new}
\end{figure}

Figure \ref{fig:barplot_ema_new}  presents the updated per-channel correlations for raw EMA targets using Wav2Vec 2.0 with BiLSTM. Tongue coordinates generally achieve higher correlations than lip coordinates, with ULz remaining one of the most challenging channels. However, vertical coordinates do not consistently outperform horizontal coordinates across all groups. 

\begin{figure}[h]
  \centering
  \includegraphics[width=0.92\linewidth]{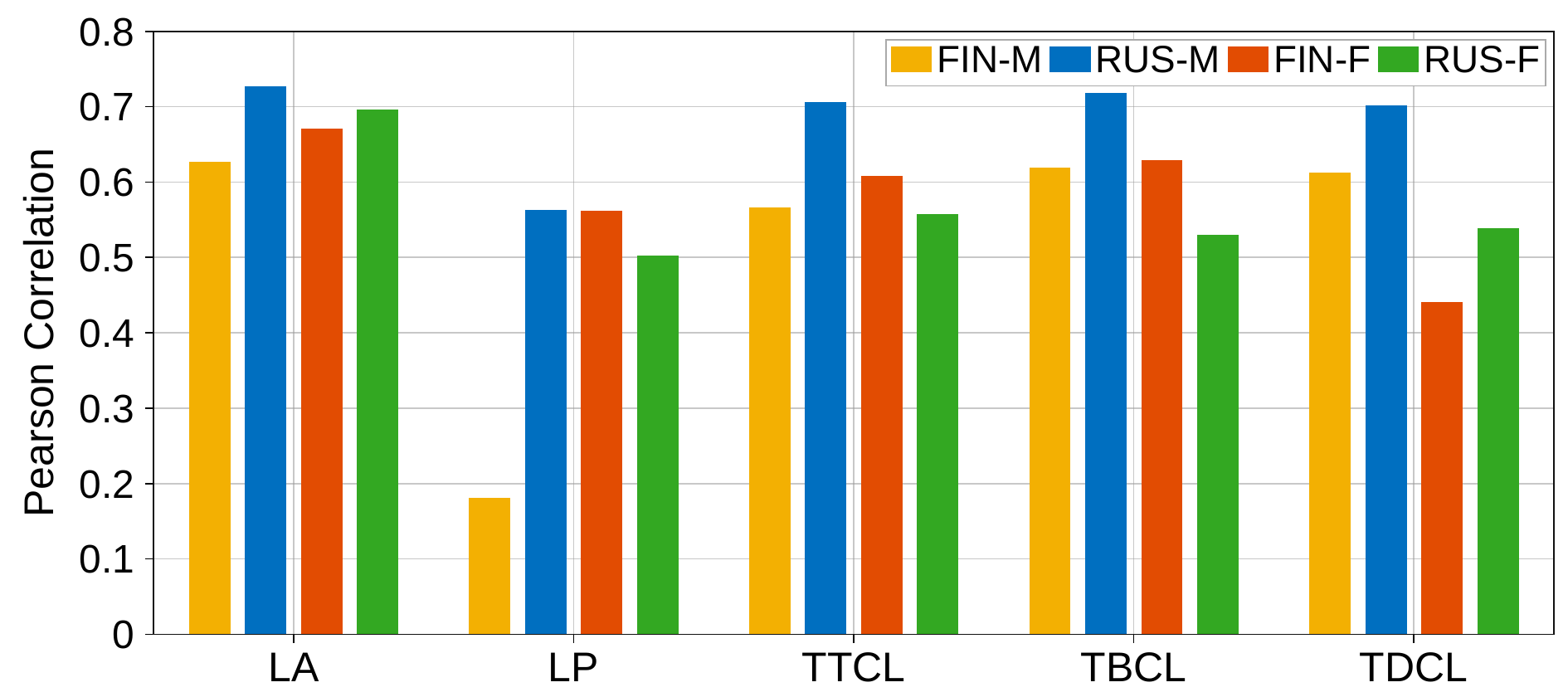}
  \caption{Per-channel Pearson correlation ($r$) for tract variable targets across four language–gender groups under in-domain LOSO evaluation using Wav2Vec 2.0 with BiLSTM.}
  \label{fig:barplot_tv_new}
\end{figure}

For the derived tract-variable (TV) targets in Figure \ref{fig:barplot_tv_new}, lip aperture (LA) is generally well predicted, whereas lip protrusion (LP) exhibits greater variability, particularly for FIN-M. The three tongue-location variables (TTCL, TBCL, and TDCL) achieve broadly comparable correlations within each group.
\subsection{Cross-gender transfer within language}
Table \ref{tab:cross_gender_SI_new} summarizes cross-gender transfer within Finnish and Russian, with the language held constant. Tongue sensors generally achieve higher correlations than lip sensors in both transfer directions. For Finnish, FIN-M $\rightarrow$ FIN-F and FIN-F $\rightarrow$ FIN-M yield comparable overall performance (r $\approx$ 0.54), while Russian transfer shows a modest directional difference (r = 0.59 for RUS-M $\rightarrow$ RUS-F and r = 0.61 for RUS-F $\rightarrow$ RUS-M). These results demonstrate that articulatory information can be transferred across genders, although the performance differences depend on the training and test groups.

\begin{table}[h]
\centering
\caption{Cross-gender speaker-independent transfer (language held constant, L1 only). Per-articulator Pearson $r$ using Wav2Vec 2.0 with BiLSTM on raw EMA targets.}
\label{tab:cross_gender_SI_new}
\setlength{\tabcolsep}{3.5pt}
\renewcommand{\arraystretch}{1.00}
\begin{tabular}{lllllll}
\toprule
Train & Test & LL & UL & TT & TB & TD \\ \hline
FIN-M & FIN-F & \cellr{0.46}{0.031} & \cellr{0.40}{0.042} & \cellr{0.65}{0.034} & \cellr{0.67}{0.019} & \cellr{0.47}{0.135} \\ \hline
FIN-F & FIN-M & \cellr{0.40}{0.061} & \cellr{0.36}{0.046} & \cellr{0.62}{0.050} & \cellr{0.67}{0.020} & \cellr{0.63}{0.018} \\ \hline
RUS-M & RUS-F & \cellr{0.53}{0.016} & \cellr{0.39}{0.108} & \cellr{0.68}{0.051} & \cellr{0.67}{0.065} & \cellr{0.68}{0.058} \\ \hline
RUS-F & RUS-M & \cellr{0.56}{0.038} & \cellr{0.43}{0.060} & \cellr{0.67}{0.026} & \cellr{0.69}{0.065} & \cellr{0.69}{0.058} \\ \bottomrule
\end{tabular}
\end{table}

\begin{figure}[h!]
  \centering
  \includegraphics[width=\linewidth]{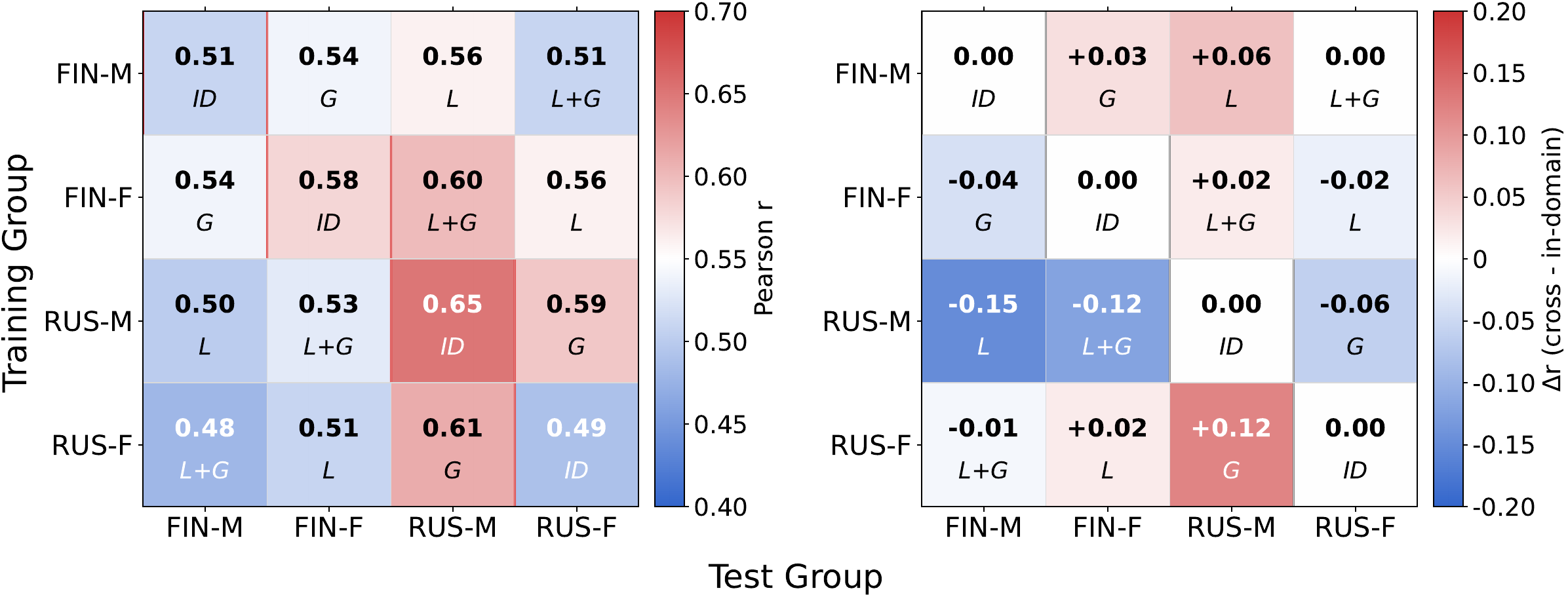}
  \caption{ Cross-domain AAI transfer (Wav2Vec 2.0, BiLSTM, raw EMA). (a) Absolute Pearson $r$ for all training–test group pairs (left). (b) Performance drop ($\Delta r$) relative to in-domain diagonal baselines (right). Cell labels indicate shift type: ID (in-domain), G (gender), L (language), L+G (combined).}
  \label{fig:Heatmap_new}
\end{figure}
\definecolor{sepgray}{HTML}{C0C0C0}
\begin{table}[h!]
\centering
\caption{Ablation over front-ends, targets, and back-ends. Pearson $r$ (std.\,dev.) on three conditions anchored to FIN-M: in-domain (ID) LOSO,
cross-gender (FIN-M$\to$FIN-F), and cross-language (FIN-M$\to$RUS-M).}
\label{tab:tab_ablation_new}
\setlength{\tabcolsep}{1.3pt}
\renewcommand{\arraystretch}{0.80}
\small
\newcommand{\rv}[2]{\makecell{#1\\[-0.5pt]{\scriptsize(#2)}}}
\newcommand{\grayrule}{%
  \arrayrulecolor{sepgray}\cmidrule(lr){3-8}\arrayrulecolor{black}}

\begin{tabular}{@{}ll ccc ccc@{}}
\toprule
\textbf{Target} & \textbf{Front-end}
& \multicolumn{3}{c}{\textbf{BiLSTM}}
& \multicolumn{3}{c}{\textbf{Attn-lite}} \\
\cmidrule(lr){3-5} \cmidrule(lr){6-8}
&
& \multicolumn{1}{c}{\textbf{LOSO}}
& \multicolumn{2}{c}{\textbf{Cross-domain}}
& \multicolumn{1}{c}{\textbf{LOSO}}
& \multicolumn{2}{c}{\textbf{Cross-domain}} \\
\cmidrule(lr){3-3} \cmidrule(lr){4-5}
\cmidrule(lr){6-6} \cmidrule(lr){7-8}
&
& ID & G & L
& ID & G & L \\
\midrule

\multirow{4}{*}{\textbf{EMA}}
  & MFCC    & \rv{0.46}{0.019} & \rv{0.44}{0.025} & \rv{0.49}{0.023}
            & \rv{0.41}{0.016} & \rv{0.36}{0.020} & \rv{0.43}{0.023} \\
\grayrule
  & Wav2Vec & \rv{0.50}{0.012} & \rv{0.53}{0.025} & \rv{0.56}{0.014}
            & \rv{0.51}{0.013} & \rv{0.55}{0.023} & \rv{0.58}{0.018} \\
\grayrule
  & XLSR-53 & \rv{0.44}{0.017} & \rv{0.50}{0.036} & \rv{0.52}{0.018}
            & \rv{0.19}{0.061} & \rv{0.30}{0.023} & \rv{0.04}{0.012} \\
\grayrule
  & MMS     & \rv{0.54}{0.011} & \rv{0.60}{0.028} & \rv{0.62}{0.021}
            & \rv{0.55}{0.014} & \rv{0.62}{0.026} & \rv{0.65}{0.021} \\
\midrule

\multirow{4}{*}{\textbf{TV}}
  & MFCC    & \rv{0.52}{0.023} & \rv{0.48}{0.025} & \rv{0.51}{0.027}
            & \rv{0.44}{0.019} & \rv{0.39}{0.021} & \rv{0.43}{0.033} \\
\grayrule
  & Wav2Vec & \rv{0.52}{0.013} & \rv{0.54}{0.022} & \rv{0.58}{0.024}
            & \rv{0.53}{0.012} & \rv{0.54}{0.021} & \rv{0.58}{0.026} \\
\grayrule
  & XLSR-53 & \rv{0.45}{0.012} & \rv{0.50}{0.030} & \rv{0.52}{0.022}
            & \rv{0.18}{0.070} & \rv{0.14}{0.027} & \rv{0.07}{0.024} \\
\grayrule
  & MMS     & \rv{0.59}{0.015} & \rv{0.61}{0.027} & \rv{0.66}{0.025}
            & \rv{0.58}{0.020} & \rv{0.62}{0.025} & \rv{0.72}{0.026} \\
\bottomrule
\end{tabular}
\end{table}
\subsection{Cross-language transfer within gender}
Figure \ref{fig:Heatmap_new} summarizes cross-domain transfer across the four language-gender groups. The updated results show that transfer performance is direction dependent, with both improvements and degradations relative to the corresponding training-group in-domain baselines. For instance, RUS-M $\rightarrow$ FIN-M yields a correlation difference of $-0.150$, whereas FIN-M $\rightarrow$ RUS-M yields $+0.059$. Similarly, RUS-F $\rightarrow$ RUS-M achieves a difference of $+0.120$. These findings indicate that the relative effects of language and gender mismatch vary across groups, and combined language-gender shifts do not consistently result in the largest performance degradation. It is worth noting that, these observations should be interpreted considering the limited and uneven speaker distribution across groups (FIN-M: 5, FIN-F: 6, RUS-M: 5, and RUS-F: 2), particularly the small RUS-F cohort.
\subsection{Ablation Study}
Table \ref{tab:tab_ablation_new} presents the updated comparison of acoustic front-ends, articulatory targets, and inversion back-ends. SSL representations generally outperform MFCCs, with MMS-300m achieving the highest overall correlations. TV targets provide performance comparable to, and often higher than, raw EMA targets. While BiLSTM remains a competitive baseline, Attn-lite matches or outperforms it in several configurations, particularly with Wav2Vec 2.0 and MMS-300m. In contrast, XLSR-53 performs relatively poorly with Attn-lite.
\section{Takeaway}
The additional 50 Hz evaluation complements the original study and supports several of its main findings, particularly the effectiveness of SSL acoustic representations and the competitive performance of interpretable articulatory targets. The updated results also provide a more nuanced understanding of cross-domain generalization and model performance across different configurations. Together, the two evaluations offer complementary perspectives on AAI using articulatory trajectories with different effective temporal resolutions and preprocessing characteristics.

\end{document}